\documentstyle[psfig,aaspp4,11pt,multicol]{article}

\lefthead{J.~M.~Miller et al.}  \righthead{\textit{XMM-Newton}
Spectroscopy of the Galactic Microquasar GRS~1758$-$258 in the Peculiar
Off/Soft State}

\received{Date}
\revised{Date}
\accepted{Date}

\journalid{vol}{date}
\articleid{1}{4}
\paperid{id}

\cpright{AAS}{1999}
\ccc{x}

\begin{document}

\title{\textit{XMM-Newton} Spectroscopy of the Galactic Microquasar
GRS~1758$-$258\\ in the Peculiar Off/Soft State\footnote{Based on
observations made with \textit{XMM-Newton}, an ESA science mission
with instruments and contributions directly funded by ESA Member
States and by NASA.}}

\author{J.~M.~Miller\altaffilmark{2},
        R.~Wijnands\altaffilmark{2,5},
	P.~M.~Rodriguez-Pascual\altaffilmark{3},\\
        P.~Ferrando\altaffilmark{4},
	B.~M.~Gaensler\altaffilmark{2,6},
	A.~Goldwurm\altaffilmark{4},
	W.~H.~G.~Lewin\altaffilmark{2},
	D.~Pooley\altaffilmark{2}
	}

\altaffiltext{2}{Center~for~Space~Research and Department~of~Physics,
        Massachusetts~Institute~of~Technology, Cambridge, MA
        02139--4307; jmm@space.mit.edu}
\altaffiltext{3}{\textit{XMM-Newton} SOC, Villafranca Satellite
        Tracking Station, PO box 50727, 28080, Madrid, Spain}
\altaffiltext{4}{Service d'Astrophysique, DSM/DAPNIA,
        CEA/Saclay, F-91191 Gif sur Yvette, Cedex, France} 
\altaffiltext{5}{\it Chandra Fellow}
\altaffiltext{6}{\it Hubble Fellow}


\authoremail{jmm@space.mit.edu}

\label{firstpage}

\begin{abstract}
We report on an \textit{XMM-Newton} Reflection Grating Spectrometer
observation of the black hole candidate and Galactic microquasar
GRS~1758$-$258.  The source entered a peculiar ``off/soft'' state in
late February, 2001, in which the spectrum softened while the X-ray
flux -- and the inferred mass accretion rate -- steadily decreased.
We find no clear evidence for emission or absorption lines in the
dispersed spectra, indicating that most of the observed soft flux is
likely from an accretion disk and not from a cool plasma.  The
accretion disk strongly dominates the spectrum in this
lower-luminosity state, and is only mildly recessed from the
marginally stable orbit.  These findings may be difficult to explain
in terms of advection-dominated accretion flow, or ``ADAF'' models.
We discuss these results within the context of ADAF models,
simultaneous two-flow models, and observed correlations between hard
X-ray flux and jet production.
\end{abstract}


\section{Introduction}
The source GRS~1758$-$258 was discovered with the
\textit{GRANAT}/SIGMA hard-X-ray/soft-$\gamma$-ray telescope (Sunyaev
et al. 1991).  It is often referred to as a ``twin'' source with
1E~1740.7$-$2942.  Both are in the vicinity of the Galactic Center
region, are strong emitters at hard-X-ray/soft-$\gamma$-ray energies,
and display relativistic jets in the radio band (Mirabel et al. 1992;
Rodriguez, Mirabel, \& Marti 1992; Mirabel 1994).  The hard spectrum
of GRS~1758$-$258 extends to 300~keV and is similar to that of
Cygnus~X-1, and on that basis it is considered a black hole candidate
(for a review of BHCs, see Tanaka \& Lewin 1995).

The soft X-ray component in this source is usually very weak when the
hard X-ray flux is at a typical strength.  The soft component was too
weak to be positively detected with \textit{ASCA} (Mereghetti et
al. 1997).  However, a soft component below 2~keV, modeled with a
power-law ($\Gamma_{\rm PL}\sim3$) was observed with \textit{ROSAT} in
1993 when the hard power-law flux was less intense (Mereghetti et
al. 1994).  Similarly, the soft X-ray component was detected with
\textit{XMM-Newton} during a short, softer episode in September, 2000
(Goldwurm et al. 2001).  In this observation, the soft component could
be fit with a simple blackbody spectrum (kT $=$ 0.32~keV).

In late February of 2001, observations with \textit{RXTE} revealed a
sharp drop in the hard ($>$10~keV) flux from GRS~1758$-$258, with no
corresponding drop in the soft flux (Smith et al. 2001a,b,c).  Smith
et al. (2001c; hereafter, SHMS) report a steady decline in the
3--25~keV flux for 50 days following 27 February 2001.  The soft
component is clearly detected with the \textit{RXTE}/PCA, and can be
fit by both simple blackbody and multicolor disk (MCD; Mitsuda et
al. 1984) blackbody models.  This state is qualitatively similar to
the ``off'' state observed by \textit{GRANAT}/SIGMA between fall 1991
and spring 1992 (Gilfanov et al. 1993), during which the hard X-ray
flux of GRS~1758$-$258 decayed beneath detection limits.  We therefore
refer to the present X-ray spectral state as the ``off/soft'' state.

We requested a \textit{XMM-Newton} target of opportunity (TOO)
observation of GRS~1758$-$258 for three principal reasons: to better
understand the nature of the soft component, to understand the
implications of the off/soft state for current accretion flow models,
and to place the off/soft state within the context of emerging
connections between spectral flux components and jets in BHCs (Fender
2001).  Herein we report the results obtained with the
\textit{XMM-Newton} Reflection Grating Spectrometer (RGS).

\section{Observations and Data Analysis}
Following our TOO request, GRS~1758$-$258 was observed for
$\sim$22~ksec starting at UT 09:48:12 on 22 March, 2001.  RGS1 and
RGS2 were operated in ``spectral$+$Q'' mode.  The EPIC-MOS cameras
were operated in ``partial window, 100$\times$100'' mode, and the
EPIC-PN camera in ``large partial window, 200$\times$384'' mode.  For
the OM, ``grism 1,'' optimized for the ultraviolet band, was used.

Accurate analysis of bright sources like GRS~1758$-$258 requires a
careful treatment of pile-up when dealing with the EPIC data.  This in
turn requires a good knowledge of the mirror response; the EPIC
calibration team is developing a description of the in-flight
calibration.  The results presented in this work are therefore
restricted to analysis of the RGS data, and results from the EPIC data
analysis will be presented in future work.  At the time of writing,
the only previous X-ray binary observations reported on with
\textit{XMM-Newton} are EXO~0748$-$676 (Bonnet-Bidaud et al. 2001;
Cottam et al. 2001) and GRS~1758$-$258 (observation in September 2000,
Goldwurm et al. 2001).  Instrumental performance and systematic
effects will become more clear as more sources with a strong soft
continuum flux are observed.

We have used the spectral data and background files produced by the
standard pipeline processing system at the \textit{XMM-Newton} Survey
Science Center (SSC) using Science Analysis Software (XMM-SAS) version
5.0.  We estimate that fewer than 1\% of the counts in the RGS spectra
are from the nearby source GX~5$-$1.

To obtain improved statistical constraints on fitted models and to
ensure the accuracy of Poisson statistics, we rebinned the
background-subtracted first-order spectra from RGS1 and RGS2 by
requiring a minimum of 20 counts per bin.  After rebinning, between
0.35--0.60~keV the flux bin errors include zero.  The flux bins
steadily increase and the errors are inconsistent with zero above
0.60~keV.  We therefore adopt 0.60~keV as the lower limit of our
fitting range.  Similarly, above 2.3~keV the flux bins have large
errors and are consistent with zero, so we adopt 2.3~keV as the upper
limit to our fitting range (the full energy range of the RGS is
0.35--2.50~keV).

Using XSPEC version 11.0 (Arnaud \& Dorman, 2000), the RGS1 and RGS2
spectra are fitted simultaneously to account for any cross-calibration
uncertainties.  An overall normalization constant is allowed to float
between the data obtained from each.  Values obtained for this
constant are approximately 0.95 for all fits.

\section{Results}

The results of our joint fits to the RGS spectra are detailed fully in
Table 1.  In each of the six models, the continuum components are
multiplied by a model for photoelectric absorption (``wabs,'' in
XSPEC).  Model 6 is a single-component model consisting of a simple
power-law.  Although statistically only marginally worse than models
dominated by a thermal component ($\chi^{2}/\nu=1.778, \nu=1251$), the
power-law index is abnormally high ($\Gamma_{\rm PL}=5.2\pm0.2$), and
inconsistent with values obtained for other BHCs regardless of
spectral state (for a review, see Tanaka \& Lewin 1995).  We therefore
conclude that models dominated by a thermal component are required to
fit the data from GRS~1758$-$258 in the off/soft state, and
concentrate on these.

In models which include thermal and power-law components, the
power-law index could not be constrained.  In order to be consistent
with models widely fit to BHCs, and especially those recently fit to
GRS~1758$-$258, we fix the power-law indices in models 1--5 to those
measured by SHMS via fits to \textit{RXTE}/PCA data obtained on 12--13
March, 2001 in the 3--25 keV band using the same models.  The fit
results for models 1--5 indicate that the power-law component is not
required in this energy range as a zero normalization is preferred.
We quote 90\% upper-limits for the 0.6--2.3~keV contribution from a
power-law component in Table 1.  In all cases, the power-law
contributes less than 2\% of the total flux in this band.  This is
consistent with \textit{Chandra} measurements in the 1--10~keV band
during the off/soft state (Heindl and Smith 2001).

Models 1 and 3 consist of MCD and simple blackbody components,
respectively, in combination with a power-law component
($\chi^{2}/\nu=1.621,~\chi^{2}/\nu=1.587$, respectively, $\nu=1252$).
The data/model ratios obtained for models 1 and 3 are nearly identical
(see Figure 1), and neither indicates the presence of any significant
line emission (indeed, line emission is not indicated regardless of
continuum model).  In part, the poor $\chi^{2}$ statistics obtained
for these models (indeed, for all models) may be due to the
approximately 5\% normalization discrepancy between RGS1 and RGS2.  A
second contributing factor may be the relatively smaller RGS effective
area above $\sim$2~keV; we included this region after rebinning to
better-constrain any power-law flux.  When the characteristics of the
RGS are more clearly known for strong continuum sources like
GRS~1758$-$258, it may be possible to improve on the fits.

Models 2 and 4 are duplicates of models 1 and 3 respectively, except
in models 2 and 4 we have fixed the neutral hydrogen column density
(N$_{H}$) to 1.5$\times10^{22}$~atoms/cm$^{2}$ as per fits to
\textit{ASCA} data of GRS~1758$-$258 reported by Mereghetti et
al. (1997).  Model 5 is like 3 and 4 in that simple blackbody and
power-law components are used to model the continuum, but in this
model N$_{H}$ is fixed to 1.74$\times10^{22}$~atoms/cm$^{2}$, as
measured via fits to \textit{XMM-Newton}/EPIC-MOS data on
GRS~1758$-$258 obtained in September, 2000 (Goldwurm et al. 2001).
There are no reasons to doubt the value of N$_{H}$ we measure due to
instrumental effects (e. g., CCD pile-up, as the RGS is a dispersive
spectrometer).  Rather, these models are fit to allow for comparisons
to the results of SHMS and Goldwurm et al. (2001) as directly as
possible.

Via the MCD model, SHMS obtain a color temperature of ${\rm
kT}=0.464\pm0.007$~keV.  Via model 1 (N$_{H}$ free), we find a color
temperature of ${\rm kT}=0.34\pm0.01$~keV.  In both measurements, the
errors are 90\% confidence errors; these color temperatures are
significantly different.  Via model 2 (N$_{H}$ fixed at SHMS value),
we find a color temperature of ${\rm kT}=0.60\pm0.01$~keV.  Again,
this is significantly different from the value measured by SHMS.
However, the fit with model 2 ($\chi^{2}/\nu=2.227, \nu=1251$) is
statistically worse than that obtained via model 1
($\chi^{2}/\nu=1.621, \nu=1252$).

SHMS report a simple blackbody temperature of ${\rm
kT}=0.395\pm0.006$~keV.  Fits with model 3 (N$_{H}$ free) indicate
${\rm kT}=0.286\pm0.007$~keV, and via model 4 (N$_{H}$ fixed at SHMS
value) ${\rm kT}=0.378^{+0.005}_{-0.004}$~keV.  The temperature
measured via model 4 is only slightly lower than that measured by
SHMS, and only marginally inconsistent at 90\% confidence.  However,
the fit obtained with model 3 is statistically preferred and the
temperature is significantly lower.

The blackbody temperature obtained via fitting with model 5, wherein
N$_{H}$ is fixed to the value measured by Goldwurm et al. (2001), is
${\rm kT}=0.332^{+0.002}_{-0.001}$~keV.  They measure ${\rm
kT}=0.32\pm0.02$~keV -- these temperatures are consistent at 90\%
confidence.  Statistically, model 3 is a marginally better fit, and
for this model the measured temperature is significantly lower.

We attempted to describe the spectrum in terms of a diffuse plasma by
fitting with the ``mekal'' and ``raymond'' models within XSPEC.  Fits
which are statistically only marginally worse than those with models
1--6 are obtained, but only if the elemental abundances are allowed to
assume values less than 0.2\% of solar values.  As this physical
scenario is extremely unlikely, we do not further consider these models.

To test for the presence of narrow emission or absorption lines, we
apply the continuum we measured via model 1 in fits to the rebinned
data to the RGS1 and RGS2 spectra at full instrumental resolution.  We
selected model 1 instead of model 3 as we regard the MCD model to be
more physical for BHCs than a simple blackbody model (the difference
in the continuum shape from model three should be very small).  In
Figure 2, we plot the 90\% confidence upper-limits on the strengths of
emission or absorption features with widths less than or equivalent to
the first-order RGS resolution.  For visual clarity, we have rebinned
the data in this plot by a factor of 10.  It is clear from this plot
that any narrow line features in the 0.6--2.3~keV range are very weak
(equivalent width $\sim$1--5~eV for most of the bandpass) when
GRS~1758$-$258 is in the off/soft state.

Finally, we examine the lightcurve obtained during this observation
for the full RGS energy range (see Figure 3).  Standard
Shakura--Sunyaev (1973) accretion disks should vary on the viscous
timescale through the disk (weeks).  In sources wherein a power-law is
observed (and a coronal volume therefore implied), significant
short-timescale flux variations (seconds--hours) are observed (see
Tanaka \& Lewin 1995).  Such rapid variations -- both aperiodic and
quasi-periodic -- can be described in terms of magnetic flaring
between the disk and corona or within the corona itself (see, e.g., Di
Matteo, Celotti, \& Fabian 1999; Merloni, Di~Matteo, \& Fabian 2000).
The lightcurve shown in Figure 3 is extremely steady, supporting an
accretion disk interpretation for the spectrum we have observed.

\section{Discussion}
Although we have observed the soft X-ray component in the off/soft
state of GRS~1758$-$258, no clear emission or absorption features have
been observed in the joint spectra of RGS1 and RGS2 (see Figures 1 and
2).  Observations of BHCs 1E~1740.7$-$2942 (Cui et al. 2001) and
SS~433 (Marshall et al. 2001) with \textit{Chandra} have found
evidence for extended regions around the X-ray sources which are too
large to be an accretion disk but too small to be a supernova remnant.
A central goal of our observation was to examine the nature of such an
extended region in GRS~1758$-$258, if one exists.  Since
optically-thin, cooling plasmas produce strong line features, it is
unlikely that the soft component we have measured is due to such a
region.  The improbable abundances required by fits with models for
diffuse plasmas support this interpretation.

The jets from GRS~1758$-$258 observed in the radio band might be
another potential source for X-ray line emission.  Tentative evidence
for lines from a jet are found in the Chandra observation of
1E~1740.7$-$2942 noted above; lines are clearly detected in the
spectra of SS~433.  The absence of lines that might be attributed to a
jet in GRS~1758$-$258 is consistent with a picture in which jets are
extinguished in the soft state in BHCs (Fender 2001).  Recently, it
has also been proposed that jets might produce the power-law component
observed in BHCs (Markoff, Falcke, \& Fender 2001).  The observed
absence of lines consistent with a jet, coupled with the diminished
strength of the power-law component in the off/soft state, may support
this model.  Stronger conclusions must await the publication of radio
data gathered during the off/soft state of GRS~1758$-$258.

The soft X-ray component we have measured is likely due primarily (but
not necessarily entirely) to an accretion disk.  This interpretation
is bolstered by the lack of significant variability in the lightcurve
obtained for this observation (see Figure 3).  We measure a disk
temperature which is broadly consistent with results obtained with the
\textit{RXTE}/PCA by SHMS on 12--13 March 2001, and with a short soft
state only five months previous to our observation (Goldwurm et
al. 2001).  From Table 1, it is clear that the temperature of the soft
component and the 0.6--2.3~keV flux are dependent upon the value of
N$_{H}$ which is used.  Based on the results of SHMS, and our finding
that a power-law component contributes less than 2\% of the flux in
the 0.6--2.3~keV range, it is possible that in the extended off/soft
state the hard power-law flux merely turns-off and reveals the
accretion disk.

The off/soft state is indicated by a steady decrease in the 3--25~keV
flux from GRS~1758$-$258, and by a far softer spectrum than is
observed when the source is in the more typical low/hard state (SHMS
2001).  Such a decrease in flux is usually interpreted in terms of a
decreasing $\dot{m}$.  Advection-dominated accretion flow (ADAF)
models adapted to BHCs (Esin, McClintock, \& Narayan 1997; Esin et
al. 1998) predict that a decrease in the mass accretion rate
($\dot{m}$) should be accompanied by spectral hardening.  The spectral
hardening may occur as the inner accretion disk is replaced by a hot,
optically-thin region which may Comptonize seed photons and produce a
power-law spectral component.  In sharp contrast, the spectrum of
GRS~1758$-$258 softens as the inferred $\dot{m}$ decreases.

If the distance to a source and its inclination are known, the MCD
model provides a measure of the temperature and inner radial extent of
the accretion disk.  In fact, the derived temperature and radius may
be distorted by the Comptonization of disk photons by a coronal volume
or a disk atmosphere.  Shimura and Takahara (1995) have reported that
a simple correction factor $f_{col}$ can account for this effect; the
effective disk temperature is given by T$_{eff}={\rm
T}_{col}/f_{col}$, and the effective radius is given by
r$_{eff}={f_{col}}^2{\rm r}_{col}$.  For BHCs, a typical value is
$f_{col}=1.7$ (see, e. g., Sobczak et al. 1999).

In Figure 4, we present constraints on the inner radial extent of the
accretion disk.  The best-fit color temperature and radius measured
with MCD models in Table 1 are used.  We assume a distance of 8.5~kpc
to GRS~1758$-$258 based on its central Galactic position and column
density.  The inclination of GRS~1758$-$258 has not yet been measured,
however the detection of jets in the radio band makes it unlikely that
the system is seen face-on.  We therefore assume an intermediate value
for the inclination angle ($\theta_{incl}=45$).  We plot constraints
on the inner radii calculated using no color correction
($f_{col}=1.0$) and the typical value for BHCs ($f_{col}=1.7$).

Given that the hard flux is diminished greatly in the off/soft state
of GRS~1758$-$258, it is likely that the Comptonization of disk
photons is not as strong an effect as in other BHCs and the color
correction may not be required.  If no color correction is applied,
for reasonable values of the black hole mass in GRS~1758$-$258 the
inner extent of the accretion disk is broadly consistent with the
marginally stable circular orbit around a Schwarzschild black hole
(see Figure 4.  This finding may be inconsistent with ADAF models of
other BHCs at relatively low implied values of $\dot{m}$ (in some
sources, however, X-ray flux and spectral states may be decoupled;
see, e. g., Homan et al. 2001; Wijnands \& Miller 2001).  If the color
correction is accurate, the inner disk extent is slightly recessed and
consistent with ADAF models of Cygnus X-1 in the low/hard state
(r$_{in}>20~{\rm R}_{\rm g}$, Esin et al. 1998).  However, the
spectrum of Cygnus~X-1 at lower implied values of $\dot{m}$ is more
consistent with an inner ADAF volume than the spectrum of
GRS~1758$-$258.  A simple ADAF interpretation might still be valid for
this source if the soft component could be attributed to an extended
cool volume, but our results indicate that this is very unlikely.  A
model in which the inner advection region is peculiarly
radiatively-inefficient might describe the off/soft state of
GRS~1758$-$258.

Smith, Heindl, and Swank (Smith et al. 2001d) have discussed the
long-term spectral variability of BHCs 1E~1740.7$-$2942,
GRS~1758$-$258, GX~339$-$4, Cygnus X-1, and Cygnus X-3 in terms of two
independent accretion flows.  As SHMS note, the off/soft state is
adequately described by this model.  The possible cooling of the
accretion disk which we measure supports this picture.  It is
interesting to note that Chen, Gehrels, and Leventhal (1994) have
proposed that 1E~1740.7$-$2942 and GRS~1758$-$258 may accrete both
through an accretion disk and via Bondi-Hoyle accretion from the
relatively high-density ISM in the Galactic center.  Observationally,
the two-flow and disk-plus-Bondi-Hoyle accretion models may be very
similar.

\section{Acknowledgments}
We wish to thank \textit{XMM-Newton} PI Fred Jansen, and the
\textit{XMM-Newton} staff for executing this target-of-opportunity
observation and their help in processing the data.  We also wish to
thank the anonymous referee for carefully reading this manuscript and
providing helpful suggestions.  RW was supported by NASA through
Chandra fellowship grants PF9-10010, which is operated by the
Smithsonian Astrophysical Observatory for NASA under contract
NAS8--39073.  BMG acknowledges the support of NASA through Hubble
Fellowship grant HST-HF-01107.01-A awarded by the Space Telescope
Science Institute, which is operated by the Association of
Universities for Research in Astronomy, Inc., for NASA under contract
NAS~5--26555.  WHGL gratefully acknowledges support from NASA.  This
research has made use of the data and resources obtained through the
HEASARC on-line service, provided by NASA-GSFC.


\begin{table}[t]
\caption{Models and Results}
\begin{small}
\begin{center}
\begin{tabular}{llllllll}
 ~ & ~ & ~ & \multicolumn{3}{l}{Multicolor Disk Blackbody Fits} & ~ & ~ \\
\tableline
Model & N$_{H}$ & kT & N$_{\rm MCD}$ & PL~index &
N$_{\rm PL}$ & L$_{0.6-2.3}$ & $\chi^{2}$/dof\\
 ~ & ($10^{22}$~cm$^{-2}$) & (keV) & ~ & ~ & ($10^{-3}$)
& ($10^{36}$~erg/s) & ~ \\
\tableline
1 & 2.28$^{+0.07}_{-0.02}$ & 0.34$^{+0.01}_{-0.01}$ &
5850$^{1900}_{1100}$ & 2.75 & $<$4.5 & $8^{+2}_{-2}$ & 2030/1252\\
2 & 1.50 & 0.60$^{+0.01}_{-0.01}$ & 180$^{+10}_{-10}$ & 2.75 & $<$0.6 & $2.6^{+0.2}_{-0.2}$ & 2786/1251\\
\tableline
 ~ & ~ & ~ & ~ & ~  & ~ & ~ & ~ \\
 ~ & ~ & ~ & \multicolumn{3}{l}{~~~~~~~~~Blackbody Fits} & ~ & ~ \\
\tableline
Model & N$_{H}$ & kT & N$_{\rm BB}$ & PL~index & N$_{\rm PL}$ & L$_{0.6-2.3}$ & $\chi^{2}$/dof\\
 ~ & ($10^{22}$~cm$^{-2}$) & (keV) & (E-3) & ~ & ($10^{-3}$) & ($10^{36}$~erg/s) & ~ \\
\tableline
3 & 2.09$^{+0.06}_{0.06}$ & 0.286$^{+0.007}_{-0.007}$ & 9.1$^{+0.9}_{-0.8}$ & 2.89 & $<$5.1 & $5.0^{+0.6}_{-0.4}$ & 1987/1252\\
4 & 1.50 & 0.378$^{+0.005}_{-0.004}$ & 4.30$^{+0.05}_{-0.05}$ & 2.89 & $<$0.6 & $2.37^{+0.03}_{-0.06}$ & 2361/1251\\
5 & 1.74 & 0.332$^{+0.002}_{-0.001}$ & 5.66$^{+0.08}_{-0.08}$ & 2.89 & $<$1.6 & $3.18^{+0.07}_{-0.04}$ & 2096/1251\\
\tableline
 ~ & ~ & ~ & ~ & ~  & ~ & ~ & ~ \\
 ~ & ~ & ~ & \multicolumn{3}{l}{~~~~~~~~~Power-law Fits} & ~ & ~\\
\tableline
Model & N$_{H}$ & -- & -- & PL~index & N$_{\rm PL}$ & L$_{0.6-2.3}$ & $\chi^{2}$/dof\\
 ~ & ($10^{22}$~cm$^{-2}$ & ~ & ~ & ~ & ~ & ($10^{36}$~erg/s) & ~ \\
\tableline
6 & 2.9$^{+0.1}_{-0.1}$ & ~ & ~ & 5.2$^{+0.2}_{-0.2}$ &
2.2$^{+0.4}_{-0.4}$ & $48^{+9}_{-8}$ & 2222/1251\\
\tableline
\end{tabular}
\vspace*{\baselineskip}~\\ \end{center} \tablecomments{Results of
fitting the background-subtracted, rebinned first-order RGS1 and RGS2
spectra jointly with standard models, in the 0.6--2.3~keV bandpass
(90\% confidence errors).  The power-law indices in models 1--5 are
fixed to the values obtained by Smith et al. (2001) using
\textit{RXTE} within the off/soft state as this component cannot be
constrained in the limited energy range of the RGS.  The power-law
component is not required in fits 1--5 and the quoted normalizations
and fluxes are upper-limits (90\% conf.).  The neutral hydrogen column
density, N$_{H}$, is allowed to float in models 1, 3, and 6.  In
models 2 and 5, N$_{H}$ is fixed to the values reported by Smith et
al. (2001) to allow for more direct comparison to those results.
Similarly, N$_{H}$ is fixed to the value reported by Goldwurm et
al. (2001) in fits to \textit{XMM-Newton} EPIC-MOS data obtained in
September, 2000.  Model 6 is a simple power-law; the obtained
$\chi^{2}$ statistic is slightly worse than the models dominated by
thermal components, but the measured index is very different than
those reported in other BHCs (see, e. g., Tanaka \& Lewin 1995).  In
fits 1--5, the power-law contribution to the 0.6--2.3~keV luminosity
is $<$2\% of the total; luminosities are calculated assuming a
distance of 8.5~kpc.  Luminosities quoted above are ``unabsorbed''
luminosities.}
\vspace{-1.0\baselineskip}
\end{small}
\end{table}

\pagebreak

\begin{figure}
\figurenum{1}
\label{fig:MCD fit}
\centerline{~\psfig{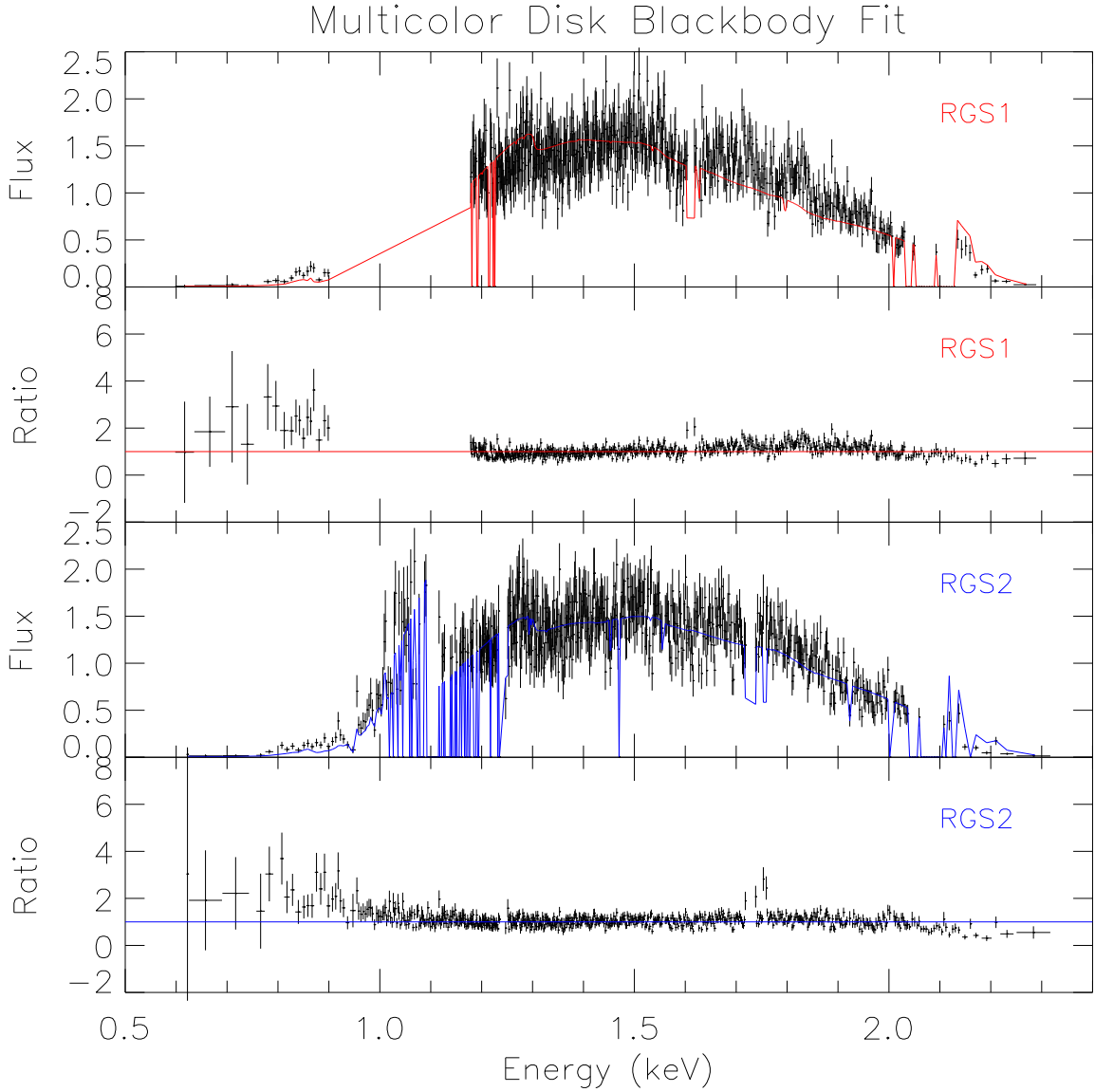}~}
\caption{Results of fitting to RGS1 and RGS2 jointly with a model
consisting of multicolor disk blackbody and power-law components.  The
flux (in units of normalized counts/cm$^{2}$/s) and data/model ratio
are shown; at top: RGS1, at bottom: RGS2.  Fits with a model
consisting of simple blackbody and power-law components, and of only a
power-law component (see Table 1) yield similar results in terms of
the fit and data/model ratio.  The large gap seen in RGS1
(0.9--1.2~keV) is due to a failed CCD in that region of the dispersed
spectrum.  Points where the model goes through flux bins consistent
with zero (in RGS1: near 1.2~keV, and 2.0--2.1~keV; in RGS2: between
1.0--1.3~keV, and 2.0--2.1~keV) are an artifact of our rebinning
(requirement of 20 counts per channel) in regions of low or rapidly
changing effective area.}
\end{figure}

\pagebreak

\begin{figure}
\figurenum{2}
\label{fig:Limits}
\centerline{~\psfig{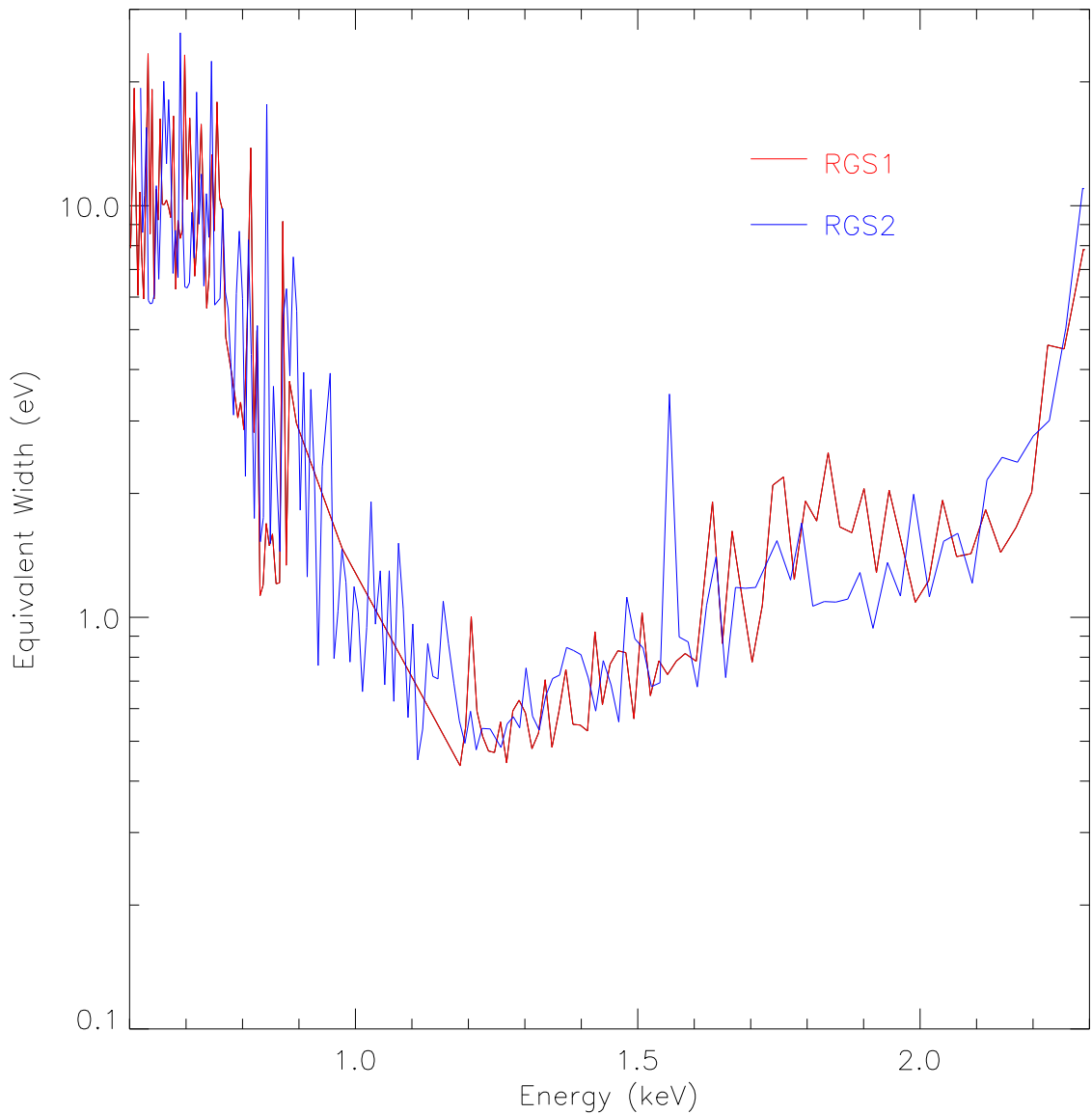}~}
\caption{Upper limits (90\% confidence) on the strength of narrow,
single-bin emission or absorption features in the RGS spectrum of
GRS~1758$-$258.  RGS1 is shown in red, and RGS2 is shown in blue.  The
limits have been rebinned by a factor of 10 for visual clarity.  These
limits are based on our best fit to the continuum with the MCD
multicolor disk blackbody model; results from a simple black body
continuum model are very similar.}
\end{figure}

\pagebreak

\begin{figure}
\figurenum{3}
\label{fig:Lightcurve}
\centerline{~\psfig{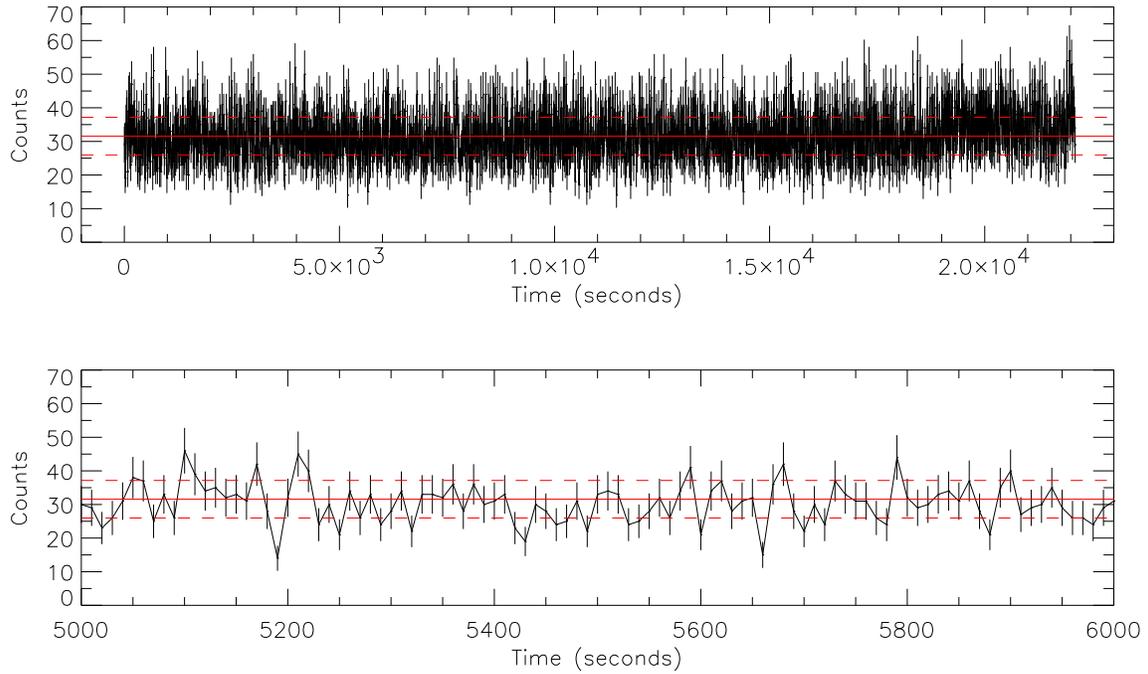}~}
\caption{The combined RGS-1 and RGS-2 first-order lightcurve (time
bins are 10 seconds, energy range: 0.3--2.5 keV).  Above, the
lightcurve for the full observation; below, a randomly selected
segment.  Plotted through the lightcurve is the mean count rate (solid
red line), and 1$\sigma$ deviations (individual error bars are
1$\sigma$).  Standard Shakura-Sunyaev (1973) accretion disks should
vary on a viscous timescale (weeks).  This lightcurve is extremely
steady, supporting our interpretation of the low-flux state of
GRS~1758$-$258 as being dominated by an accretion disk.}
\end{figure}

\pagebreak

\begin{figure}
\figurenum{4}
\label{fig:Radius}
\centerline{~\psfig{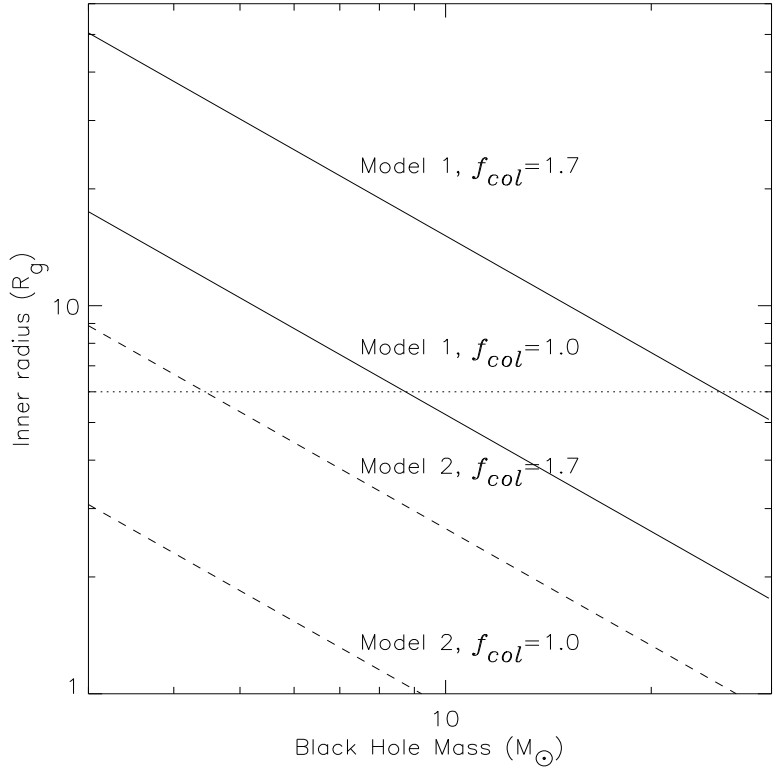}~}
\caption{Inner accretion disk radii derived via the multicolor disk
blackbody model, assuming an intermediate inclination ($\theta_{\rm
incl}=45^{\circ}$) and a distance of 8.5~kpc.  Fits with model 1
(N$_{H}$ free) yield inner radii of 77~km and 220~km, for
$f_{col}=1.0$ and $f_{col}=1.7$, respectively.  Fits with model 2
(N$_{H}$ fixed) yield inner radii of 14~km and 39~km, for
$f_{col}=1.0$ and $f_{col}=1.7$, respectively.  The marginally stable
circular orbit around a Schwarzschild black hole is 6~R$_{\rm g}$;
this radius is indicated with a dotted line.}
\end{figure}

\end{document}